\documentclass[aps,prd,showpacs,nobibnotes,twocolumn,preprintnumbers,nobalancelastpage,superscriptaddress]{revtex4-1}

\usepackage{natbib}
\setcitestyle{numbers}
\usepackage{amssymb} 
\usepackage{amsmath} 
\usepackage{graphicx}
\usepackage{latexsym} 
\usepackage{dcolumn}
\usepackage{bm}
\usepackage{color}
\input{epsf} 
\newcommand{\be}{\begin{equation}}
\newcommand{\ee}{\end{equation}} 
\newcommand{\bea}{\begin{eqnarray}}
\newcommand{\eea}{\end{eqnarray}} 
\newcommand{\ba}{\begin{array}}
\newcommand{\ea}{\end{array}} 
\newcommand{\no}{\nonumber}
\newcommand{\bfi}{\begin{figure} \epsfxsize=8cm \epsffile}
\newcommand{\bfig}{\begin{figure*} \epsfxsize=15cm \epsffile}
\newcommand{\efi}{\end{figure}} 
\newcommand{\efig}{\end{figure*}}
\newcommand{\bi}{\begin{itemize}} 
\newcommand{\ei}{\end{itemize}}
\newcommand{\mpch}{h^{-1} {\rm Mpc}} 
\newcommand{\hmpc}{h {\rm Mpc}^{-1}}

\newcommand{\dif}{\mathrm{d}}

\newcommand{\bmx}{{\bm{x}}}
\newcommand{\bmr}{{\bm{r}}}
\newcommand{\bmk}{{\bm{k}}}
\newcommand{\bmv}{{\bm{v}}}

\newcommand{\bmve}{{\bm{v_E}}}
\newcommand{\bmvb}{{\bm{v_B}}}

\newcommand{\bibmnras}{Mon. Not. R. Astron. Soc. }
\newcommand{\bibapj}{Astrophys. J. }

\newcommand{\bibjcap}{J. Cosmol. Astropart. Phys. }
\newcommand{\bibprd}{Phys. Rev. D }
\newcommand{\bibprl}{Phys. Rev. Lett. }
\newcommand{\bibphysrep}{Physics Reports }

\newcommand{\bibaap}{Astron. Astrophys. }
\newcommand{\bibnat}{Nature (London) }

\begin{document}

\title{Kriging interpolating cosmic velocity field. II: Taking
  anisotropies and multi-streaming into account}

\author{Yu Yu}
\email{yuyu22@shao.ac.cn}
\affiliation{Key laboratory for research in galaxies and cosmology,
Shanghai Astronomical Observatory, Chinese Academy of Sciences, 80
Nandan Road, Shanghai 200030, China}
\author{Jun Zhang}
\affiliation{Center for Astronomy and Astrophysics, Department of Physics and Astronomy,
Shanghai Jiao Tong University, 200240, Shanghai}
\affiliation{IFSA Collaborative Innovation Center, Shanghai Jiao Tong University, Shanghai 200240, China}
\author{Yipeng Jing}
\affiliation{Center for Astronomy and Astrophysics, Department of Physics and Astronomy,
Shanghai Jiao Tong University, 200240, Shanghai}
\affiliation{IFSA Collaborative Innovation Center, Shanghai Jiao Tong University, Shanghai 200240, China}
\author{Pengjie Zhang}
\email{zhangpj@sjtu.edu.cn}
\affiliation{Center for Astronomy and Astrophysics, Department of Physics and Astronomy,
Shanghai Jiao Tong University, 200240, Shanghai}
\affiliation{IFSA Collaborative Innovation Center, Shanghai Jiao Tong University, Shanghai 200240, China}
\affiliation{Key laboratory for research in galaxies and cosmology,
Shanghai Astronomical Observatory, Chinese Academy of Sciences, 80
Nandan Road, Shanghai 200030, China}

\begin{abstract}
Measuring the volume-weighted peculiar velocity statistics from
inhomogeneously and sparsely distributed galaxies/halos, by
existing velocity assignment methods,  suffers from a significant
sampling artifact. As an alternative,  the Kriging interpolation based
on Gaussian processes was introduced and  evaluated [Y. Yu, J. Zhang, Y. Jing,
and P. Zhang, Phys. Rev. D 92, 083527 (2015)]. Unfortunately,  the
most straightforward application of Kriging does not perform better than the existing
methods in the literature. In this work, we investigate two physically motivated
extensions. The first takes into account of the  anisotropic velocity
correlations. The second introduces the nugget effect, on account of
multi-streaming of the velocity field.   We find that the performance
is indeed improved.  For sparsely sampled data
[$n_P\lesssim 6\times 10^{-3}(\mpch)^{-3}$] where the sampling
artifact is the most severe, the improvement is significant and is
two-fold: 1) The scale of reliable measurement of the
velocity power spectrum is extended by a factor $\sim 1.6$,
 and 2) the dependence on the velocity correlation prior is weakened by a factor of $\sim 2$.
We conclude that such extensions are desirable for accurate velocity
assignment by  Kriging. 
\end{abstract}

\maketitle

%%%%%%%%%%%%%%%%%%%%%%%%%
%%%%%%%%%%%%%%%%%%%%%%%%%

\section{Introduction}
\label{sec:introduction}
Peculiar velocity of galaxies and other tracers of the large-scale
structure such as free electrons and 21 cm emitting neutral hydrogen atoms probes the structure
growth rate of the Universe. It is therefore a valuable probe of dark
matter, dark energy and gravity at cosmological scales  (e.g.,
Refs. \cite{Kaiser87,Peacock01,Linder03,zhangpj07,Guzzo08,Jain08,
  wangyun08,Reyes10,limiao11,Clifton12,Reid12, 
  Tojeiro12,Weinberg13,Joyce15,Koyama15}). It also offers the
possibility of probing horizon-scale inhomogeneities of the Universe and
therefore tests the external inflation
\cite{ZhangStebbins11,ZhangJohnson15}. In different circumstances, the
measured velocity statistics can have different weighting. For
example, the kinetic Sunyaev-Zel'dovich effect is 
proportional to the gas momentum, which is the gas density-weighted
velocity.  On the other hand, one can infer
the volume-weighted velocity power spectrum from redshift space
distortions (RSD), by comparing the measured RSD power spectrum with
the theoretical modelling. In this approach, the RSD theory models the volume-weighted velocity
statistics (e.g., Ref. \cite{Zhang13}). 

The density-weighted velocity statistics suffers from the
problem of density bias, which is hard to calculate from the first
principle. In contrast,  the volume-weighted velocity statistics is
free of this problem.  Therefore, for the purpose of precision
cosmology, it  is more desirable than the density-weighted  velocity
statistics. For this aspect, RSD cosmology requires accurate
understanding of the volume-weighted velocity statistics. Due to the
nonlinear evolution of the large-scale structure, the most
robust way of understanding the volume-weighted velocity statistics is
through cosmological simulations. Through them, we can measure the
velocity statistics of simulation particles, halos and mock galaxies. 

However, accurate measurement of the volume-weighted velocity in
simulations is nontrivial,  due to the very fact that we only sample
the velocity field where there are simulation particles (e.g.,
Refs. \cite{Bernardeau96,Schaap00,Zheng13,Koda14,zhangpj15,zhengyi15a,Jennings15}). Since
the particle distribution is inhomogeneous and since  it
is correlated with the signal (velocity) that we aim to measure, the
sampling of the velocity field is biased. This sampling artifact increases
with decreasing the sample number density $n_P$ (e.g., \cite{zhangpj15}). It is already detectable
for $n_P\sim 1 {\rm Mpc}^{-3}$ \cite{Zheng13}. For sparse samples
such as massive halos, the problem is much more severe. For example,
when  $n_P\sim 10^{-4} {\rm Mpc}^{-3}$,  the induced error in the velocity
power spectrum increases to $\mathcal{O}(10)\%$,  even at scales as large as
$k=0.1h/$Mpc \cite{zhengyi15a}. 

To solve/alleviate this long-standing problem, several methods of
velocity assignment have been proposed: 
\begin{itemize}
\item[(1)] The Voronoi tessellation (VT)
method \cite{Bernardeau96} is a zeroth-order interpolation scheme. 
One constructs the Voronoi tessellation from a set of nodes (i.e., particles/halos).
The velocity inside a tessellation element (i.e., the Voronoi
polyhedral) is approximated as the velocity of the only particle enclosed. 
One then obtains the velocity on regular grids by smoothing this
space-filling velocity field. 
\item[(2)] The Delaunay tessellation (DT) method \cite{Schaap00} is a linear interpolation scheme.
One first constructs the Delaunay tessellation, which is the dual of Voronoi tessellation.
One then approximates the velocity gradient inside a Delaunay tessellation element (i.e., the
Delaunay tetrahedron) as a constant,  determined
by the velocities of the four vertices.  The two steps construct the
velocity field at all positions. One can then apply a smoothing onto
the interpolated velocity field. 
\item[(3)] In the works by \citet{Zheng13} and \citet{Koda14}, the
nearest-particle (NP) method was proposed and applied. 
It assigns the velocity of a regular grid point as the velocity of the
nearest particle to it. It is
essentially the VT method without smoothing.  
\end{itemize}
\citet{zhangpj15} studied
the sampling artifact theoretically and verified the theoretical
modelling through simulations \cite{zhengyi15a}. One finding is that the polynomial
interpolation  has to be at  least quadratic, in order to be free of the
leading-order sampling artifact. Therefore, sampling artifacts of the above velocity assignment
methods are all severe for sparse samples,  consistent with the simulation tests of the NP method
\cite{zhengyi15a} and the DT method \cite{Jennings15}.

This motivates us to try alternative methods.  In our previous work
\cite{yuyu15b} (hereafter paper I), the Kriging method was introduced
and tested. The Kriging interpolation, originally used in
geostatistics, is an application of Gaussian processes. It assumes the
Gaussian distribution for the field to be interpolated. With a set of
positions with known field values (data) and adopting priors on the
spatial correlation function of the field,  the posterior distribution of the field 
value at any other spatial location can be predicted.  Kriging then
assigns the peak value of such posterior distribution as the field value at
the given position. Therefore,  Kriging is a maximum likelihood
estimator. Furthermore, one can prove that this peak value is the
weighted linear combination of field values of  nearby positions [Eq. \ref{eqn:wy}], and the resulting error
dispersion is minimal (a detailed explanation can be found in paper
I). Therefore it is also a minimal variance linear
estimator.  Actually, this provides the most straightforward viewpoint of
Kriging and leads to the most direct derivation of the Kriging
formula presented in paper I and in the current paper. 

However, despite the above desirable properties,  we found that the
versions of Kriging investigated in paper I  do not  perform better
than the NP method for reconstructing the 3D velocity field on regular
grids.  In Kriging, 2 degrees of freedom in Kriging can be explored to improve its performance: 
1) One can choose the set of positions with known values to perform the interpolation. 
For example, one can use the $n_k$ particles nearest the given point, 
and in the limiting case of $n_k=1$, it reduces to the NP method.  
Since the particle distribution is inhomogeneous,
the spatial range of interpolation varies from position to
position \footnote{This differs from another minimal variance linear estimator,
the Wiener filter. The Wiener filter has a position independent
window function }. Paper I has investigated the cases of
$n_k=[10,200]$.  2) Another degree of freedom is the priors on the spatial correlation function of
the field.  Paper I fixed the velocity correlation function as that
predicted by the Lambda cold dark matter ($\Lambda$CDM) cosmology while trying two different
values of $\Omega_m$. When the sample density is low, the performance
of Kriging is sensitive to the adopted priors (paper I). This
motivates us to adopt priors of better physical motivations. 

One motivation is that the physical velocity correlation function $\xi_{\alpha\beta}({\bf
  r})$ 
between the $\alpha$th component at position ${\bf r}_1$ and the $\beta$th
component at position ${\bf r}_2$ is anisotropic. Namely $\xi_{\alpha\beta}({\bf
  r})\equiv \langle v_\alpha({\bf r}_1) v_\beta({\bf r}_2)\rangle$ depends on not only the amplitude of the separation vector ${\bf
  r}\equiv {\bf r}_2-{\bf r}_1$ but also on its direction $\hat{r}\equiv {\bf r}/r$.  In fact, it can be decomposed into two isotropic correlation functions $\psi_\perp(r)$ and $\psi_\parallel(r)$ \cite{Peebles80},
\be
\label{eqn:anisotropyxi}
%\label{eqn:xi2d}
\xi_{\alpha\beta}(\bmr)=\psi_\perp(r)\delta_{\alpha\beta}+[\psi_\parallel(r)-\psi_\perp(r)]\hat{r}_\alpha\hat{r}_\beta\ .
\ee
Here, $\alpha,\beta=x,y,z$ indicates the three Cartesian
axes. $\delta_{\alpha\beta}$ is the Kronecker delta
function. Figure \ref{fig:psi} shows $\psi_{\perp,\parallel}(r)$ in a
$\Lambda$CDM universe. The two show visible inequality, meaning
significant anisotropy in $\xi_{\alpha\beta}({\bf r})$, as shown in
Fig. \ref{fig:xi2d}.  Paper I 
ignored such anisotropy and adopted the averaged correlation function
$\xi_{vv}\equiv \langle {\bf v}({\bf r}_1)\cdot{\bf
  v}({\bf r}_2)\rangle=(\psi_{\parallel}+2\psi_{\perp})/3$.  Since the
anisotropy is $\mathcal{O}(10)\%$ at $r=10 \mpch$ and can be even
larger at larger $r$, this is expected to result in significant error
in the reconstructed velocity field. 

Other physics missing in paper I are the multistreaming of the velocity
field. In the late stage of structure formation, shell crossing develops,
and the velocity field is no longer single valued. This means that,
even when two particles share the same
position, they may not share the same velocity. In terms of the
correlation function, 
\bea
\langle
v_\alpha^2\rangle-\xi_{\alpha\alpha}(r\rightarrow 0)\neq 0\ .
\eea
In the language of Kriging, this corresponds to a variogram nonzero
at $r=0$, namely, the nugget effect. Paper I ignored this complexity. In the current paper, we
will take it into account and test whether it can improve the Kriging
performance. 

For clarity, we call the Kriging investigated in paper I ``Kriging
I'' and the version considering anisotropies and multi-streaming of
the velocity field ``Kriging II''.  This paper is organized as follows.
In Sec. \ref{sec:kriging}, we briefly introduce the general idea of
Kriging,  Kriging I, and Kriging II. In Sec. \ref{sec:simulation}, we
briefly describe the simulation and velocity statistics we use. In
Sec. \ref{sec:results},  we present the performance test of Kriging II.
We conclude and discuss in Sec. \ref{sec:conclusion}.

%%%%%%%%%%%%%%%%%%%%%%%%%
%%%%%%%%%%%%%%%%%%%%%%%%%

\section{Kriging Interpolation as a velocity assignment method}
\label{sec:kriging}

\subsection{General idea of Kriging}
\label{sec:generalkriging}

We want to estimate the field value $y_*$ at a given position $\bmx_*$ based
on the $n_k$ points at $\bmx_i$ with observed values $y_i=y(\bmx_i)$.
Kriging predicts $y_*$ as a weighted linear combination of $y_i$,
\be
\label{eqn:wy}
y_*=\sum_i W_i y_i\ .
\ee
Here, the weighting $W_i$ depends on the position $\bmx_i$, but not on
$y_i$. It satisfied the unbiased condition $\sum_i W_i=1$. 
By minimizing the rms error with respect to the data,
the weighting $\bm{W}$ is determined.
Therefore, Kriging relies on the {\it variogram} of the field, 
\bea
\gamma(\bmr)&=&\frac{1}{2}\langle [y(\bmx+\bmr)-y(\bmx)]^2 \rangle \ , \no
\\
&=& \langle y^2\rangle-\langle y(\bmx+\bmr)y(\bmx)\rangle\ .
\eea

With the variogram prior specified, the weighting $\bm{W}$ is solved from the following Kriging system:
\be
\left[\begin{array}{c} \bm{W} \\ \mu \end{array}\right]
=\left[\begin{array}{cc} \bm{\gamma}_{ij} & \bm{1} \\ \bm{1^T} & 0 \end{array}\right]^{-1}
\left[\begin{array}{c} \bm{\gamma}_{i*} \\ 1 \end{array}\right]\ .
\label{eqn:kriging}
\ee
Here, $\bm{\gamma}_{ij}$ is the matrix containing the variogram values between observed points,
 with $\gamma_{ij}\equiv \gamma(\bmx_i-\bmx_j)$.
It has a dimension of $n_k\times n_k$.
Unity vector $\bm{1^T}$ is set for the unbiased condition $\sum_i W_i=1$.
The vector $\bm{\gamma}_{i*}\equiv \gamma(\bmx_i-\bmx_*)$.
$\mu$ is the Laplacian multiplier.
We refer the readers to the Appendix of paper I for the derivation of
the Kriging system.

\subsection{Kriging I}
To proceed, we need to adopt a variogram. Kriging I of paper I makes two
simplifications. Kriging I adopts the direction-averaged variogram
from linear theory and therefore neglects the anisotropic in the
velocity field.  The adopted variogram is given by
$\gamma(r)=1-\xi_{vv}(r)/\xi_{vv}(0)$. Here, $\xi_{vv}(r)\equiv \langle {\bf
  v}({\bf r}_1)\cdot {\bf v}({\bf r}_2)\rangle$, with no dependence on
the direction of ${\bf r}$. 

This choice of variogram automatically satisfies
$\gamma(r\rightarrow 0)\rightarrow 0$. Under such a condition, $y({\bf
  x}_*\rightarrow \bmx_i)\rightarrow y_i$. Namely, Kriging 
honors the observed values at observed positions. The condition
$\gamma(r\rightarrow 0)\rightarrow 0$  is satisfied as long as the field
is single valued.

However, in reality, the velocity correlation function is anisotropic,
and the velocity field is multi-valued in some regions such as
halos. Therefore, we propose Kriging II to improve the velocity assignment.

%%%%%%%%%%%%%%%%%%%%
\bfi{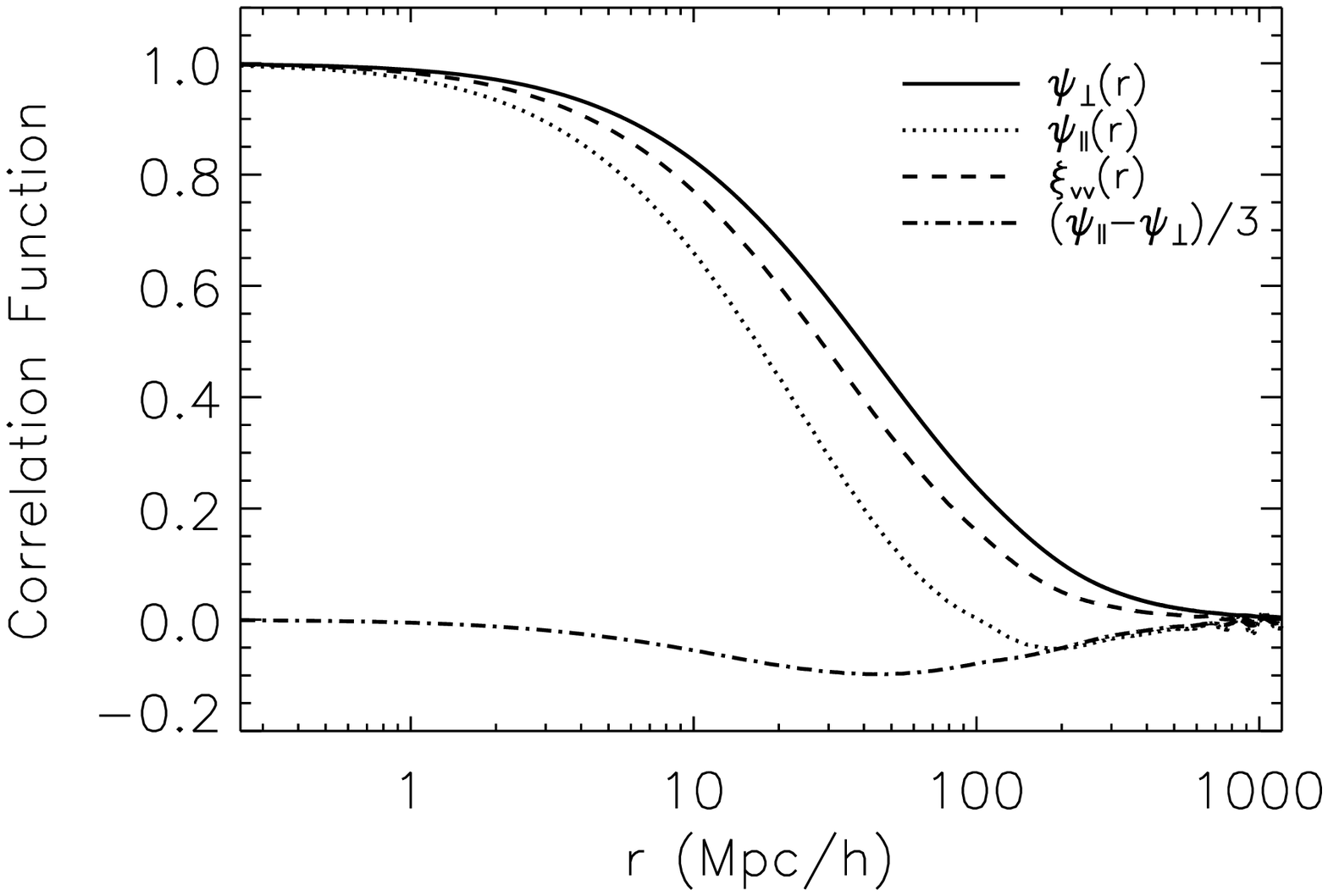}
\caption{The theoretically predicted two velocity correlation functions $\psi_\perp(r)$ (solid line) and $\psi_\parallel(r)$ (dotted line)
that form the anisotropic velocity correlation function [Eq. (\ref{eqn:anisotropyxi})]. 
The direction-averaged velocity correlation function adopted in Kriging I is presented by the dashed line.
The direction-averaged velocity correlation between different velocity components is presented by the dot-dashed line.
\label{fig:psi}}
\efi
%%%%%%%%%%%%%%%%%%%%

%%%%%%%%%%%%%%%%%%%%
\bfig{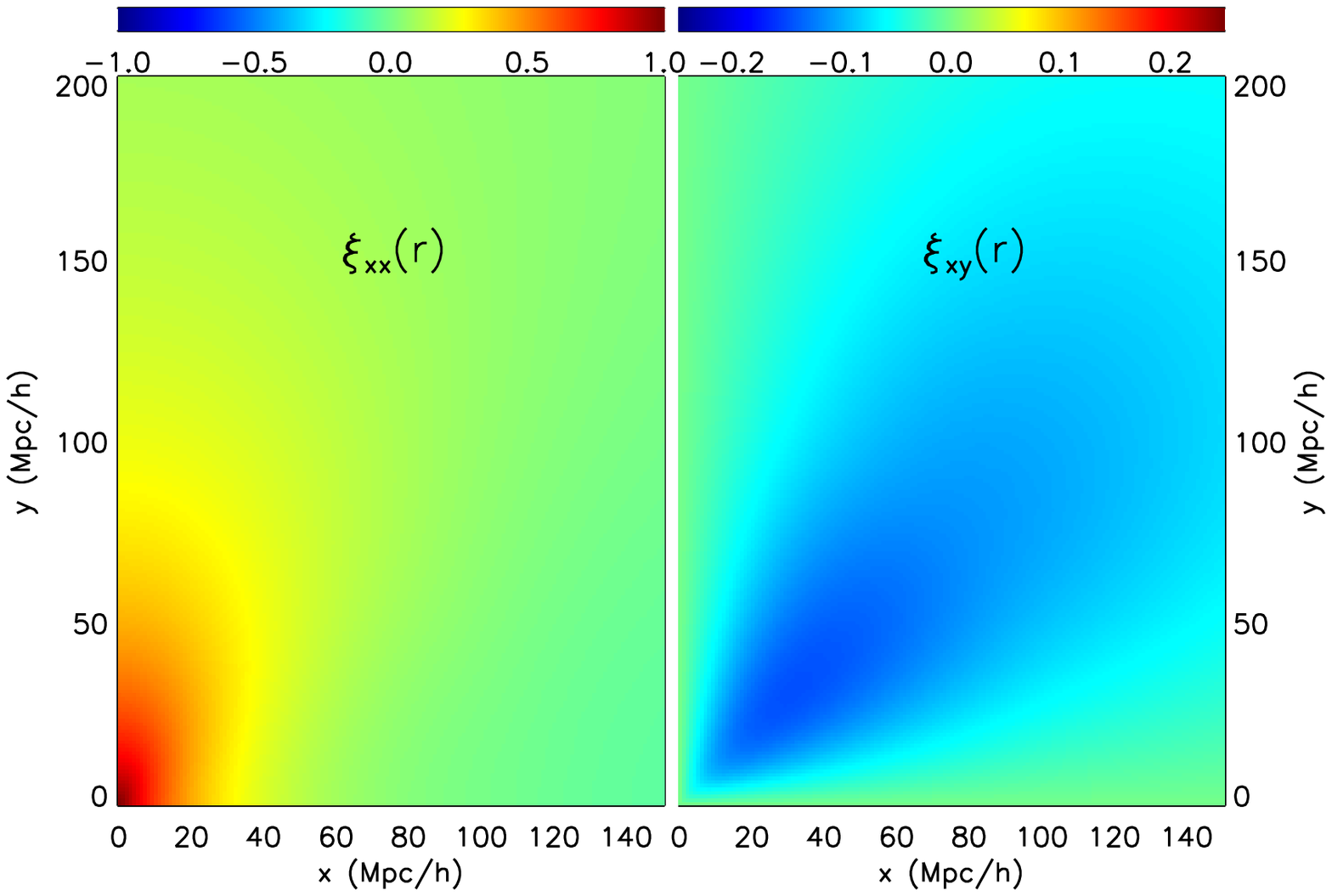}
\caption{The theoretically predicted normalized velocity correlation function between the same components [e.g., $\xi_{xx}(\bmr)$] and 
between different components [e.g., $\xi_{xy}(\bmr)$] is presented in the left and right panels, respectively. 
Strong anisotropy can be seen in the two velocity correlation functions.
For the $x$ velocity component, the correlation $\xi_{xx}(\bmr)$ is stronger in the perpendicular $y$ direction and relatively weaker in the parallel $x$ direction.
Notice that the two panels adopt different color scales to better present the anisotropy.
\label{fig:xi2d}}
\efig
%%%%%%%%%%%%%%%%%%%%

\subsection{Kriging II}
\label{sec:ekriging}
The anisotropic velocity correlation function  has the general form of
Eq. (\ref{eqn:anisotropyxi}), with two functions
$\psi_{\parallel,\perp}(r)$. At large scales of interest, the velocity
field is curl free. We then have the textbook result \cite{Peebles80}
\be
\label{eqn:psiperp}
\psi_\perp(r)=3H^2\int \frac{\dif k}{k} \frac{\Delta^2_{\theta\theta}}{k^2}\left[\frac{\sin(kr)}{(kr)^3}-\frac{\cos(kr)}{(kr)^2}\right]\ ,
\ee
\be
\label{eqn:psipara}
\psi_\parallel(r)=3H^2\int \frac{\dif k}{k} \frac{\Delta^2_{\theta\theta}}{k^2}\left[\frac{\sin(kr)}{kr}-2\frac{\sin(kr)}{(kr)^3}+2\frac{\cos(kr)}{(kr)^2}\right]\ .
\ee
Here, $H$ is the Hubble parameter, and
$\Delta^2_{\theta\theta}(k)=k^3P_{\theta\theta}(k)/2\pi^2$ is the
velocity divergence power spectrum. 
$\psi_\parallel$ and $\psi_\perp$ are not independent.  The two
satisfy  the following consistency relation:
\be
\psi_\parallel(r)=\frac{\dif (r\psi_\perp(r))}{\dif r}\ .
\ee
Theoretically predicted (normalized) $\psi_\perp$ and $\psi_\parallel$
based on $\Lambda$CDM cosmology are presented in
Fig. \ref{fig:psi}. The two differ significantly.  It results in  the
anisotropic velocity correlation function $\xi_{xx}(\bmr)$  and
$\xi_{xy}(\bmr)$ (Fig. \ref{fig:xi2d}).  The shown anisotropy implies that, 
to interpolate one specific velocity component, the optimal weighting
should differ for particles at the same $r$ but different $\hat{r}$. 
It also implies that for a fixed position, the weighting for different velocity components is different.

In principle, we can construct $v_\alpha$  from $v_\beta$ with $\beta\neq
\alpha$, for the reason that $\xi_{\alpha\beta}\neq 0$. However, since the
correlation when $\alpha\neq \beta$ is in general 
much weaker than the case of $\alpha=\beta$ (Fig. \ref{fig:xi2d}), the reconstruction is the
most accurate for the case of $\alpha=\beta$. Hereafter, we will only
consider this case; namely, we interpolate $v_\alpha$ at some given
positions to obtain $v_\alpha$ at other positions.  The corresponding
variogram, normalized at $r=0$,  is 
\bea
\label{eqn:gammaab}
\gamma_{\alpha\alpha}(\bmr)&\equiv&
1-\xi_{\alpha\alpha}(\bmr)/\xi_{\alpha\alpha}(0)\ ,\no\\
&=&\gamma_\perp(r)+[\gamma_\parallel(r)-\gamma_\perp(r)]\hat{r}_\alpha^2\ .
\eea
Here, $\gamma_\perp(r)\equiv 1-\psi_\perp(r)/\psi_\perp(0)$, and
$\gamma_\parallel(r)\equiv 1-\psi_\parallel(r)/\psi_\parallel(0)$. 

An implicit assumption in the above results is that the velocity field
has no multi-streaming, namely, is single valued. So, it automatically
satisfies $\gamma_{\alpha\alpha}(r\rightarrow 0)=0$. Although this is a desirable
property for some cases,  it is inappropriate when there  exist measurement errors or the true field
itself is multi-valued. Since in this paper we restrict to
simulations, we can neglect the measurement error.  But in
simulations, the velocity field is indeed multi-valued (multi-streaming), due
to the nonlinear evolution induced shell crossing. For example, in massive clusters, particle
motions are randomized, and we no longer have a single-valued velocity
field. This results in $\gamma(r\rightarrow 0)\neq 0$ (the {\it
  nugget} effect). To model this effect, we introduce a dimensionless free parameter
$\gamma_{\rm nug}$ and modify the variogram in Eq. (\ref{eqn:gammaab}) as 
\be
\label{eqn:KII}
\gamma_{\alpha\alpha}\rightarrow
\left[1-\frac{\xi_{\alpha\alpha}(\bmr)}{\xi_{\alpha\alpha}(0)}\right]+\gamma_{\rm
    nug}\ .
\ee
It can be rewritten as $\gamma_{\alpha\alpha}=(1+\gamma_{\rm
  nug})(1-\xi_{\alpha\alpha}(\bmr)/(\xi_{\alpha\alpha}(0)(1+\gamma_{\rm
nug}))$. 
Since Kriging does not depend on the overall amplitude of the
variogram, the modification [Eq. (\ref{eqn:KII})] is equivalent to 
$\xi_{\alpha\alpha}(0)\rightarrow \xi_{\alpha\alpha}(0)(1+\gamma_{\rm
  nug})$.

We then Kriging interpolate the velocity field for each velocity
component with the corresponding variogram ($\alpha=x,y,z$), 
\be
v_{\alpha,*}=\sum_i W_{\alpha,i}v_{\alpha,i}\ .
\ee
Due to the anisotropic velocity correlation, ${\bf W}_x\neq {\bf
  W}_y\neq {\bf W}_z$. This is different from the case of isotropic
velocity correlation, in which ${\bf W}_x= {\bf
  W}_y= {\bf W}_z$.  Thus, in Kriging II, we need to solve the Kriging system for each individual component.
The computational expense is tripled compared to Kriging I.

%%%%%%%%%%%%%%%%%%%%%%%%%
\begin{figure}
\epsfxsize=8cm
\epsffile{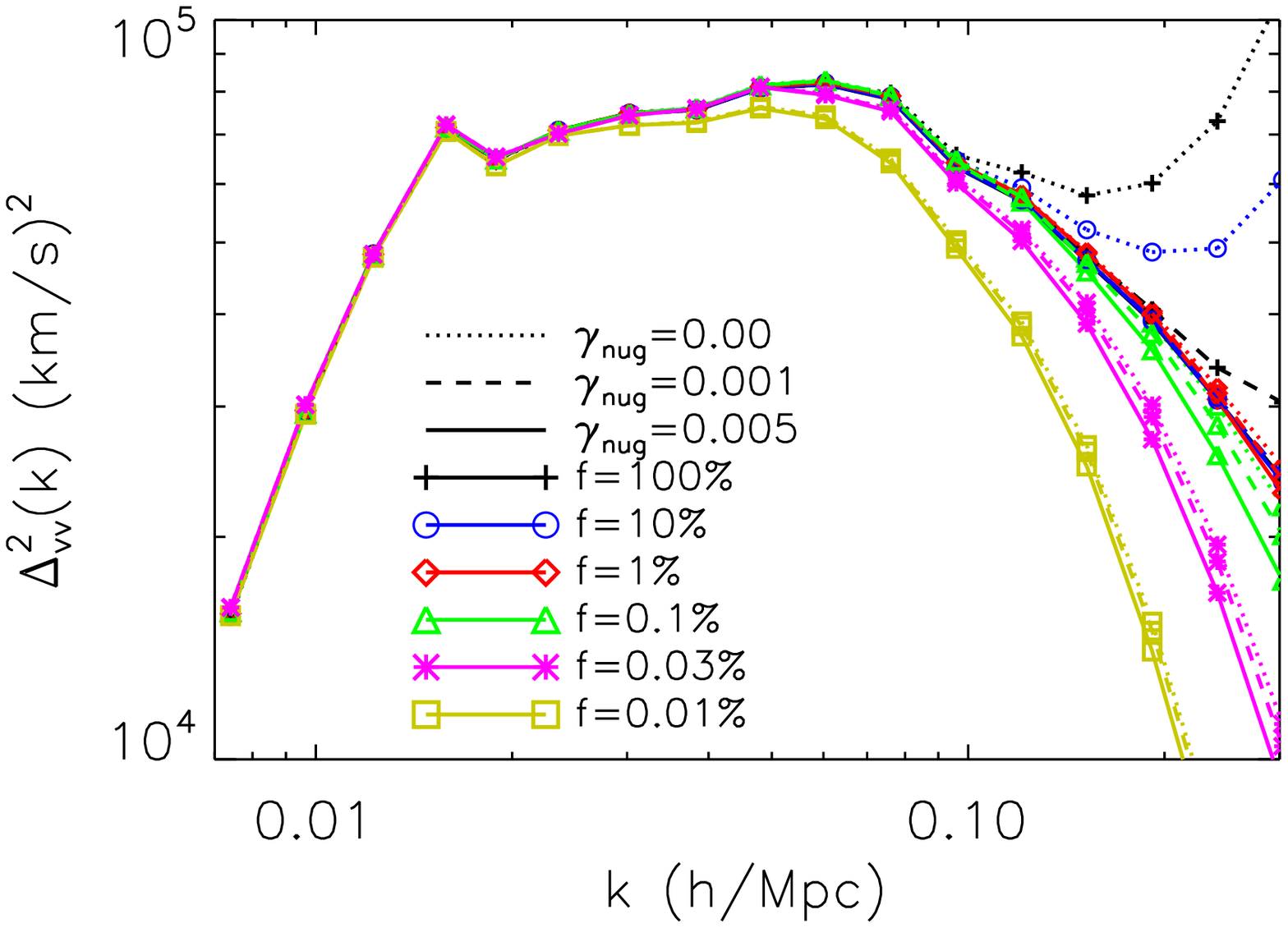}
\caption{E-mode velocity power spectrum from the Kriging method in which we arbitrarily choose the nugget values $\gamma_{\rm nug}=0$, $0.001$, and $0.005$ (dotted, dashed, and solid lines, respectively).
For $f=100\%$ and $10\%$, the estimated power without the nugget effect is boosted up in small scales, due to the instability in the Kriging system.
Taking the nugget effect into consideration alleviates this problem but smooths the power a little in small scales.
}
\label{fig:nugget}
\end{figure}
%%%%%%%%%%%%%%%%%%%%%%%%%

%%%%%%%%%%%%%%%%%%%%%%%%%
%%%%%%%%%%%%%%%%%%%%%%%%%

\section{Simulation Specification}
\label{sec:simulation}

\subsection{Simulation}

We test the performance of Kriging II using the same simulation data
used in paper I to test Kriging I.
The simulation is run by a particle-particle-particle-mesh code adopting the standard flat $\Lambda$CDM model,
 with cosmological parameters $\Omega_m=0.268$, $\Omega_\Lambda=0.732$, $\sigma_8=0.85$, $n_s=1$, and $h=0.71$ (see Ref.  \cite{jingyp07}).
The dark matter particle number is $1024^3$, and the box size is
$1200\mpch$.

We adopt the same trick as in Kriging I to test Kriging II. The sampling
artifact vanishes in the limit of infinite particle number
density. As the consequence, the full
simulation sample with $1024^3$ simulation particles and the NP \cite{Zheng13}
velocity assignment method robustly predicts the E-mode
velocity power spectrum at scales of interest ($k\lesssim 0.1\hmpc$; 
see paper I). Therefore, we treat such a result as the reference.  We then
construct low number density samples by randomly selecting a fraction
of $f$ ($f=10\%, 1\%, 0.1\%, 0.03\%, 0.01\%$) particles from the
simulation.  The number density of these samples scales with $f$ as
\be
n_P=\frac{1024^3}{1200^3}f\ (\mpch)^{-3}=0.62f\ (\mpch)^{-3} \ .
\ee
These samples have decreasing number density and therefore an increasing
sampling artifact. In particular,  future
spectroscopic surveys such as  Euclid
\footnote{http://sci.esa.int/euclid/}, DESI \footnote{http://desi.lbl.gov/},
and PFS \footnote{http://pfs.ipmu.jp/intro.html} will have a galaxy
number density comparable to that of $f=0.1\%$.  Thus, the performance
of Kriging II in this range of $f$ is of particular interest.
Comparing to paper I, we add $f=0.03\%$ in this work, since Kriging I
degrades significantly between $f=0.1\%$ and $0.01\%$ 
(see paper I).  

We construct the velocity field on $256^3$ uniform grid by Kriging II,
using the $n_k$ simulation particles nearest to a given grid point. 
The corresponding grid size is $4.7\mpch$, safe for the scales of
interest ($k\sim 0.1\hmpc$). In principle, we may use as many simulation
particles as possible for Kriging. But in reality,  we only need a
very limited fraction of them. One reason is that only nearby
particles have a non-negligible contribution in the 
Kriging interpolation. Distant particles have very weak correlation with the velocity of the given grid point.
 They have no contribution in Kriging.? Another reason is that in dense
regions, only a small amount of particles is needed for accurate
interpolation. Following studies in paper I, $n_k=200$ is sufficient
for the purpose of our study. Thus, in this work, we fix $n_k=200$ for all the cases.

\subsection{Statistics}
\label{sec:statistics}
We restrict the test to the velocity E-mode power spectrum, which
contains most of the cosmological information.  Any vector field can
be decomposed into an E-mode (gradient) and a B-mode (curl)
component. We decompose the velocity field on the uniform grid points
into an E-mode and a B-mode component by using a fast Fourier transform,
\bea
\bmve(\bmk)=\frac{(\bmv(\bmk)\cdot\bmk)\bmk}{k^2}\ ,\\
\bmvb(\bmk)=\bmv(\bmk)-\bmve(\bmk)\ .
\eea
Traditionally, we present the measured E-mode velocity power spectrum
\be
P_{\bmv\bmv}(k)=\langle\bmve\bmve^*\rangle\
\ee
in the form of $\Delta^2_{\bmv\bmv}(k)=k^3P_{\bmv\bmv}(k)/2\pi^2$.

The velocity power spectra of the simulation samples with $f<1$ should
be statistically identical to that of the full sample ($f=1$), if free of the
sampling artifact. Furthermore, since these samples share the same
cosmic variance as the full sample, their relative difference is free
of cosmic variance. Therefore, the difference in the velocity power
spectra of these samples should be interpreted as a sampling
artifact. We perform the above test on three simulation snapshots, $z=0$, $1.07$, and $1.87$,
 to quantify the redshift dependence of Kriging II. We also compare
 Kriging I and Kriging II  to see whether there is improvement by considering the anisotropy in the variogram.

%%%%%%%%%%%%%%%%%%%%%%%%%
\begin{figure*}
\epsfxsize=8cm
\epsffile{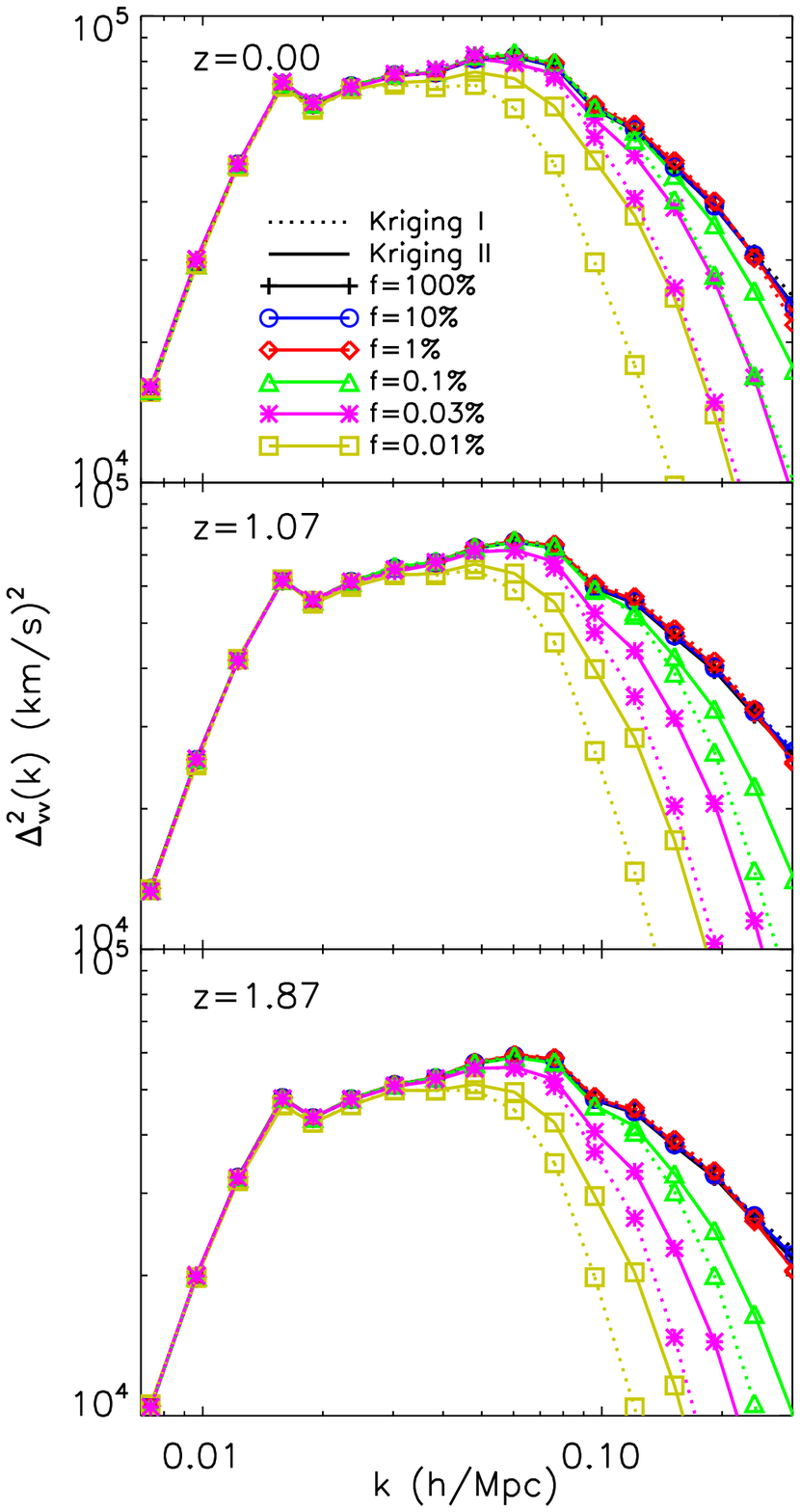}
\epsfxsize=8cm
\epsffile{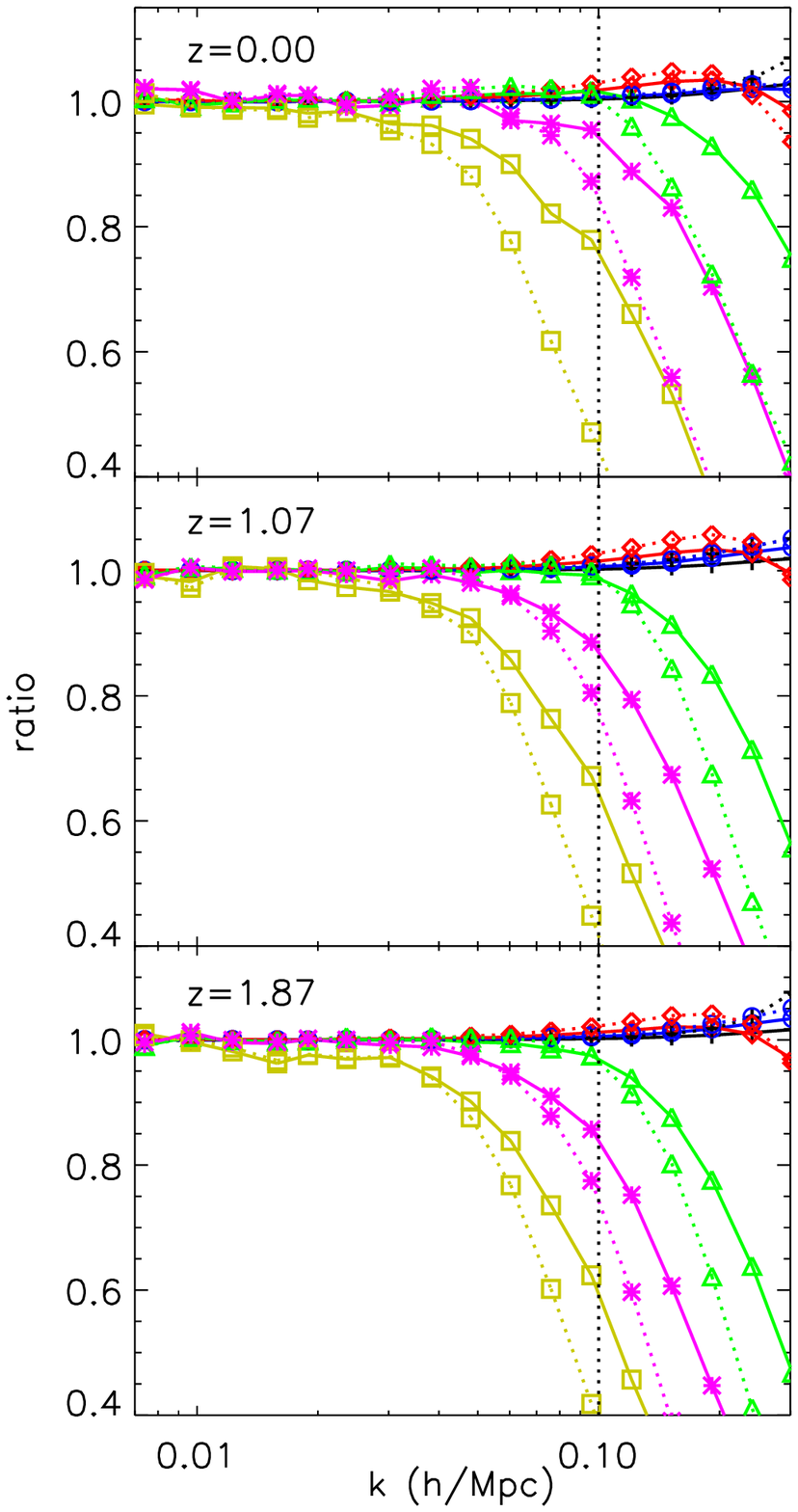}
\caption{E-mode velocity power spectrum from the Kriging method is presented in the left panel.
The power from Kriging II is presented by solid lines,
 while the result of Kriging I is presented by dotted lines.
Obvious improvement is observed.
From top to bottom, the result for $z=0, 1.07, 1.87$ is presented. 
The right panel presents the ratio to the reference cases.
The vertical dotted line indicates the most concerned scale $k=0.1\hmpc$.
\label{fig:veaniso}}
\end{figure*}
%%%%%%%%%%%%%%%%%%%%%%%%%

%%%%%%%%%%%%%%%%%%%%
\bfi{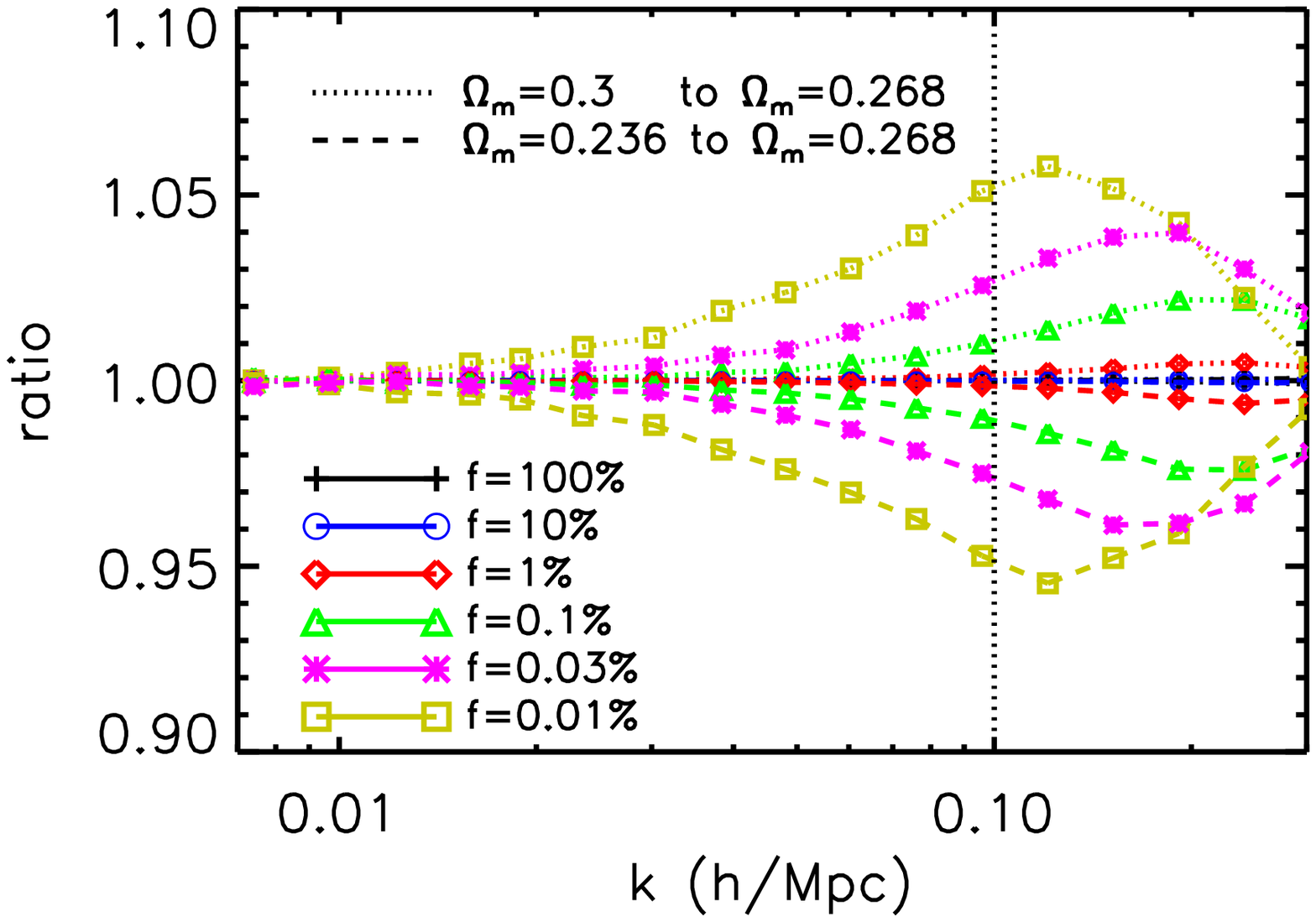}
\caption{E-mode velocity power spectrum deviations induced by adopting two inconsistent variograms predicted from a $\Omega_m=0.3$ and $\Omega_m=0.236$ flat universe.
The vertical dotted line indicates the most concerned scale $k=0.1\hmpc$.
For $f=0.1\%$, $0.03\%$, and $0.01\%$, the deviation is controlled at $1\%$, $3\%$ and $5\%$ level at $k=0.1\hmpc$.
Compared to the result in paper I, the sensitivity on the variogram prior is weakened by considering the more reasonable variogram in Kriging II.
\label{fig:cosmo}}
\efi
%%%%%%%%%%%%%%%%%%%%

%%%%%%%%%%%%%%%%%%%%%%%%%
\begin{figure}
\epsfxsize=8cm
\epsffile{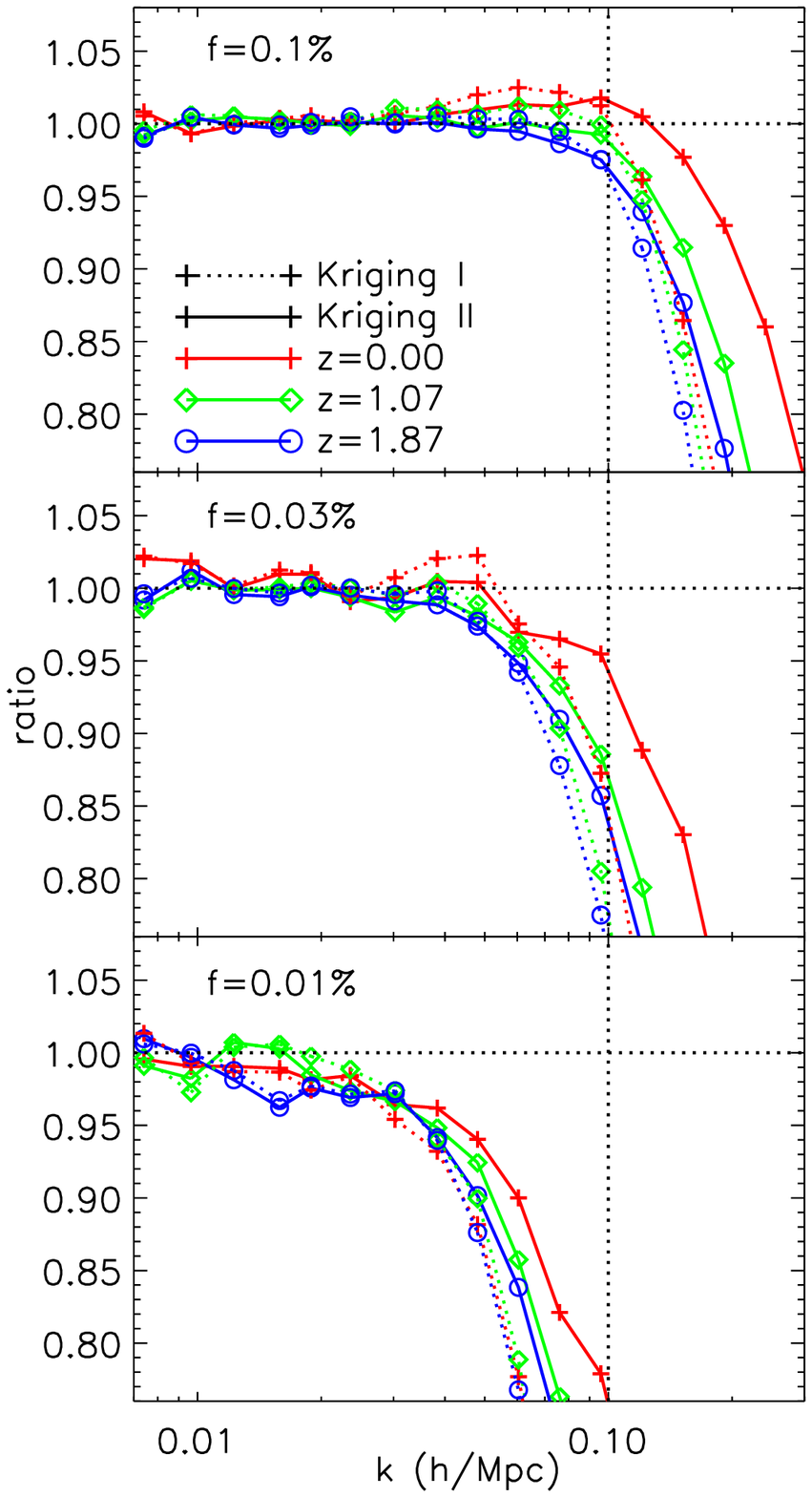}
\caption{We present the redshift dependence of the Kriging performance for $f=0.1\%$, $0.03\%$, and $0.01\%$ from top to bottom, respectively.
For a given $f$, the ratio of the E-mode velocity power spectrum to the reference case
is presented for both Kriging I (dotted line) and Kriging II (solid line).
Both Kriging I and Kriging II perform better at lower redshift.
\label{fig:veanisoratio}}
\end{figure}
%%%%%%%%%%%%%%%%%%%%%%%%%

%%%%%%%%%%%%%%%%%%%%%%%%%
%%%%%%%%%%%%%%%%%%%%%%%%%

\section{Performance of Kriging II}
\label{sec:results}

%\subsection{Anisotropy in the variogram}
For the variogram prior, we adopt the linear velocity divergence power spectrum
$\Delta^2_{\theta\theta}$ generated by the {CLASS} code
\footnote{http://class-code.net/}. This is then used to calculate
$\psi_\perp(r)$ and $\psi_\parallel(r)$ following
Eqs. (\ref{eqn:psiperp}) and (\ref{eqn:psipara}) and the result is shown in
Fig. \ref{fig:psi}. $\xi_{\alpha\beta}({\bf r})$ is calculated using
Eq. (\ref{eqn:anisotropyxi}), and the results of $\xi_{xx}$ and $\xi_{xy}$ are
shown in Fig. \ref{fig:xi2d}.  The correlation functions and therefore
the variograms are clearly anisotropic. 

The velocity power spectra obtained by Kriging II at $z=0$, with the
variogram specified above  are
presented by the dotted line in Fig. \ref{fig:nugget}, for various $f$.
We found that for the highest two sampling fraction cases (i.e., $f=100\%$ and $10\%$), 
the resulting E-mode power spectra boost up abnormally at small
scales.  In these two cases, the Kriging system is found to be
unstable for a large fraction of grid points. For these grid points,
the resulting weighting does satisfy $\sum_i W_i=1$.  However, most of the
 $W_i$ are abnormally large.
This implies that the Kriging system is almost
degenerate for some velocity component(s) on these grid points. 
We also check the condition number for the matrix in Eq. (\ref{eqn:kriging}),
 and it also implies that the matrix is ill conditioned.

The fact that this behavior only appears for the densest samples implies
the possible cause of multi-streaming. Kriging II, with a variogram
depending on the directions between the given grid point at ${\bf
  x}_{\rm grid}$ and nearby particles at ${\bf x}_{\rm grid}+{\bf r}$,
may lead to inconsistent solutions.  For example,   $v_x$ information
of nearby particles at positions ${\bf x}_{\rm  grid}+(r,0,0)$
contributes significantly to the reconstruction of $v_x({\bf x}_{\rm
  grid})$. Here, $r$ is much smaller than the velocity correlation
length. Particles at ${\bf x}_{\rm  grid}+(0,r,0)$ also have
comparable contribution. But due to multi-streaming,  particles at
${\bf x}_{\rm  grid}+(r,0,0)$ and ${\bf x}_{\rm  grid}+(0,r,0)$ can
have very different $v_x$. Such differences can confuse Kriging II,
leading to the observed degenerate Kriging system and the abnormal
behavior found above.

To alleviate this problem, we  include the nugget effect following Eq.
(\ref{eqn:KII}), and try different values of the nugget parameter
$\gamma_{\rm nug}$. Overall, it causes suppression of small
scale power, as expected if it is caused by multi-streaming. For $\gamma_{\rm nug}=0.001$, it solves
the problem for $f=10\%$ (Fig. \ref{fig:nugget}) , but the abnormal
boost of power at the small scale still remains for the $f=100\%$ case. For $\gamma_{\rm
  nug}=0.005$,  it fully solves the problem for all the cases we
consider. In contrast, it has a negligible effect on the cases of $f<1\%$
and scales of $k\sim 0.1\hmpc$. This also supports our speculation that
the nugget effect is related to multi-streaming. For low number density samples, 
there are essentially no close particle pairs.   
The above small scale multistreaming induced problem will not happen.
 Hereafter, we will fix $\gamma_{\rm
  nug}=0.005$.

\subsection{Improvements by Kriging II}
The  $z=0$  E-mode velocity power spectrum  obtained by Kriging II is presented in the top
panel of Fig. \ref{fig:veaniso} by solid lines and is compared to
that obtained by Kriging I (dotted lines).  For low number density
samples, the improvement of Kriging II over Kriging I is
significant.  The severe underestimation of small-scale power by Kriging I is
significantly alleviated for $f=0.1\%$, $0.03\%$, and $0.01\%$. For
example, for $f=0.01\%$, the $\sim 55\%$ underestimation at
$k=0.1\hmpc$ by Kriging I is reduced to $\sim 20\%$ by Kriging II.
As a consequence, it expands the scale of reliable measurement  by a factor of $\sim
1.6$. 

Another problem found in paper I is that the velocity statistics
obtained by Kriging I  has non-negligible dependence on the
variogram, and this dependence is more severe for low number density
samples.  For example, if we change the variogram from that of the fiducial
$\Omega_m=0.268$ flat $\Lambda$CDM cosmology to a $\Omega_m=0.3$ flat
$\Lambda$CDM cosmology, the velocity power spectrum at $k=0.1\hmpc$
changes by $2\%$ for $f=0.1\%$ and by $8\%$ for $f=0.01\%$ (Fig. 5, paper
I). Kriging II still has such unsatisfactory dependence on variogram
prior (Fig. \ref{fig:cosmo}). Nevertheless, it is much weaker, and the
changes are $1\%$ for $f=0.1\%$ and $5\%$ for $f=0.01\%$.  The dependence on the variogram prior is
one major limiting factor of the cosmological application of
Kriging. Therefore, although this problem is not fully solved, the improvement by
Kriging II is a notable step forward.

\subsection{Redshift dependence}
The improvement of Kriging II over Kriging I is also observed at other
redshifts. Figure \ref{fig:veanisoratio} shows the comparison at two
arbitrarily chosen redshifts, $z=1.07$ and $1.87$. In particular, for
low number density samples, Kriging II produces a larger velocity power
spectrum, closer to the correct result. 

An interesting behavior of Kriging is that its performance is worse
at higher redshift. This is likely related to the redshift dependences
of two characteristic scales. As in paper I,  we define the
characteristic length for the velocity field, 
\be
L_v=\left( \frac{\langle(\nabla\cdot\bmv)^2\rangle}{\langle\bmv^2\rangle} \right)^{-\frac{1}{2}}\ .
\ee
We then measure it from the full simulation sample with a negligible
sampling artifact.  This characteristic length describes the length on
which the velocity field varies significantly.  
The measured  $L_v$ is $6.64$, $6.00$, and $5.83$ $\mpch$ for $z=0$, $1.07$, $1.87$, respectively.
Another scale is the mean particle separation,
\be
L_P=\frac{1200}{1024} f^{-1/3}\ \mpch\ . 
\ee
The $L_P$ values for decreasing $f$ from $f=100\%$, $10\%$, $1\%$, $0.1\%$, $0.03\%$, and $0.01\%$ are $1.17$, $2.52$, $5.44$, $11.7$, and $25.3$ $\mpch$, respectively.

To reliably sample the velocity field, we should have $L_P\lesssim
L_v$. This condition is satisfied when $f\gtrsim 0.1\%$, but violated
when $f\lesssim 0.1\%$ for all three redshifts we investigated. This explains that for all redshifts, the Kriging method suddenly fails for $f\lesssim 0.1\%$.

%%%%%%%%%%%%%%%%%%%%%%%%%
%%%%%%%%%%%%%%%%%%%%%%%%%

\section{Conclusion and Discussion}
\label{sec:conclusion}
Velocity assignment is a crucial ingredient of understanding
the statistics of a cosmological velocity field. Motivated by the
existence of a sampling artifact in existing velocity assignment
methods, we introduced the Kriging method (Kriging I) in paper I. The current
paper investigates physically motivated modifications of Kriging I. The modifications in Kriging
II  are an anisotropic variogram taking the anisotropy of velocity
correlation into account and the nugget effect taking the
multi-streaming into account. At $k=0.1\hmpc$ of interest, we find
significant improvement for low number density samples. 

Unfortunately, even the improved Kriging II still biases the velocity
power spectrum at $k=0.1\hmpc$ by $\mathcal{O}(5)\%$ for samples of number
density $\sim 10^{-4} {\rm Mpc}^{-3}$. Such a systematic error is larger than
the requirement from the stage IV dark energy projects. Therefore,
further improvements are still required. We have tried other
modifications  in the framework of Kriging but found no success. As a
reference for possible future studies of Kriging as a velocity
assignment method, we show one example. Kriging I and II interpolate
one velocity component of nearby particles (e.g., $v_x$ of particles)
to obtain the same velocity component (e.g., $v_x$ on the grid). Since
$\xi_{\alpha\beta}\neq 0$ even when $\alpha\neq \beta$ (e.g., right
panel of Fig. \ref{fig:xi2d}), we can Kriging interpolate
$v_y$ of nearby particles to obtain $v_x$ on the grid. Such a version of Kriging is called the \textit{Kolmogorov-Wiener prediction}.
Thus, any data that correlate with the target data can be used to help the prediction.
This is called \textit{co-Kriging}. This uses more
information of particle velocities and can in principle help the
velocity assignment.   We have tried co-Kriging, by extending the
variogram matrix to include all $\alpha$, $\beta$ pairs. 
However, we found that this process is extremely unstable.
For most of the grid points, the Kriging system is degenerate, unless
extremely low $n_k \sim 5$ is adopted. Furthermore, even for such a low $n_k$, the smoothing effect of Kriging dominates, leading to a large suppression in the resulting power, as reported in paper I.
Therefore, co-Kriging does not help for a practical situation. 

We have also tried methods other than Kriging, NP, and
tessellations. For example,  we are looking for methods that can avoid
the interpolation onto a regular grid.  We have tried the nonuniform fast Fourier transform
method \footnote{http://www.cims.nyu.edu/cmcl/nufft/nufft.html}. 
Finally, we found that it  essentially provides the  mass-weighted
velocity. Although it was a failed attempt, we present this negative result
as a useful reference for future studies of better velocity assignment
methods. 

%%%%%%%%%%%%%%%%%%%%%%%%%
%%%%%%%%%%%%%%%%%%%%%%%%%

\section*{Acknowledgments}

This work was supported by the National Science Foundation of China (Grants No. 11403071, No. 11320101002, No. 11433001, No. 11621303, No. 11653003), National Basic Research Program of China (973 Programs No. 2013CB834900 and No. 2015CB857001),  Key Laboratory for Particle Physics, Astrophysics and Cosmology, Ministry of Education, and Shanghai Key Laboratory for Particle Physics and Cosmology (SKLPPC).   J.Z. is supported by the national Thousand Talents Program for distinguished young scholars and the T.D. Lee Scholarship from the High Energy Physics Center of Peking University. This work made use of the High Performance Computing Resource in the Core Facility for Advanced Research Computing at Shanghai Astronomical Observatory.

%%%%%%%%%%%%%%%%%%%%%%%%%
%%%%%%%%%%%%%%%%%%%%%%%%%

\bibliographystyle{apsrev4-1}
\bibliography{mybib_ZPJ}

\end{document}